\documentclass[numreferences,options]{kluwer}
\usepackage{amssymb}
\usepackage{bbm}
\newtheorem{Theorem}{Theorem}[section]
\newtheorem{Proposition}[Theorem]{Proposition}
\newtheorem{Lemma}[Theorem]{Lemma}

\newtheorem{Definition}[Theorem]{Definition}
\newproof{Remark}{Remark}
\newproof{Example}{Example}
\newproof{Proof}{Proof}
\newcommand{\CC}{{\mathbb{C}}}
\newcommand{\II}{{\mathbbm{1}}}

\newcommand{\ZZ}{{\mathbb{Z}}}
\newcommand{\bfk}{{\bf k}}
\newcommand{\bfkred}{\red{\bfk}}
\newcommand{\bfm}{{\bf m}}

\newcommand{\calB}{{\cal B}}
\newcommand{\calG}{{\cal G}}

\newcommand{\calM}{{\cal M}}
\newcommand{\con}{{\cal A}^k}
\newcommand{\conp}{\calB^k}

\newcommand{\diff}{{\rm d}}

\newcommand{\facs}[2]{\mbox{$#1$$;$$#2$}} 
\newcommand{\gau}{{\cal G}^{k+1}}
\renewcommand{\gcd}[1]{\langle#1\rangle}
\newcommand{\gref}[1]{{\rm (\ref{#1})}}

\newcommand{\Howe}{{\mbox{\rm Howe}}}

\newcommand{\mod}{{\mbox{\rm mod}}}

\newcommand{\orb}{{\cal M}^k}

\newcommand{\otG}{\type(\conp/G)}
\newcommand{\otgau}{\type(\con/\gau)}
\newcommand{\pgau}{{\cal G}_\ast^{k+1}}

\newcommand{\red}[1]{#1^\circ}

\newcommand{\rmC}{{\rm C}}
\newcommand{\rmd}{{\rm d}}

\newcommand{\rmeven}{{\rm even}}

\newcommand{\rmK}{{\rm K}}

\newcommand{\rmM}{{\rm M}}

\newcommand{\rmP}{{\rm P}}
\newcommand{\rmS}{{\rm S}}
\newcommand{\rmSO}{{\rm SO}}
\newcommand{\rmSU}{{\rm SU}}
\newcommand{\rmT}{{\rm T}}
\newcommand{\rmU}{{\rm U}}
\newcommand{\rref}[1]{{\rm \ref{#1}}}

\newcommand{\Sub}{{\rm Sub}}
\newcommand{\SUJ}{{\rmSU(J)}}
\newcommand{\SUJc}{{\rmSU(J^c)}}

\newcommand{\twk}{\tilde{k}}

\newcommand{\type}{{\rm Type}}

\newcommand{\punkt}[1]{\put(#1){\circle*{0.06}}}

\newcommand{\linie}[3]{\put(#1){\line(#2){#3}}}

\newcommand{\whole}[3]{\put(#1){\punkt{0,0}\put(0.05,0.1){\makebox(-0.1,
   -0.2)[#2]{\tiny $#3$}}}}

\newcommand{\plene}[3]{\put(#1){\punkt{0,0}\linie{0.1,0}{1,0}{0.8} 
\put(0.05,0.1){\makebox(-0.1,-0.2)[#2]{\tiny $#3$}}}}
\newcommand{\plenz}[3]{\put(#1){\punkt{0,0}\linie{0.1,0}{1,0}{1.8}
\put(0.05,0.1){\makebox(-0.1,-0.2)[#2]{\tiny $#3$}}}}

\newcommand{\plzee}[3]{\put(#1){\punkt{0,0}\linie{0.0894,0.0447}{2,1}{0.821}
\put(0.05,0.1){\makebox(-0.1,-0.2)[#2]{\tiny $#3$}}}}

\newcommand{\plzmee}[3]{\put(#1){\punkt{0,0}\linie{0.0894,-0.0447}{2,-1}{
0.821}\put(0.05,0.1){\makebox(-0.1,-0.2)[#2]{\tiny $#3$}}}}

\begin{document}
\begin{article}
\begin{opening}
\title{On a Certain Stratification of the Gauge Orbit Space}
%
\author{G. \surname{Rudolph}}
\author{M. \surname{Schmidt}\email{matthias.schmidt@itp.uni-leipzig.de}}
\institute{Institute for Theoretical Physics, University of Leipzig,
Augustusplatz 10, 04109 Leipzig, Germany}
%
\begin{abstract}
For a principal $\rmSU(n)$-bundle over a compact manifold of dimension
$2,3,4$, we determine the orbit types of the action
of the gauge group on the space of connections modulo pointed local gauge
transformations. We find that they are given by Howe subgroups of
$\rmSU(n)$ for which a certain characteristic equation is solvable. 
Depending on the base manifold, this equation leads to a linear, bilinear, 
or quadratic Diophantine equation.
\end{abstract}
\classification{2000 MSC}{53C05, 53C80}
\end{opening}
\section{Introduction}
%
%
%
Despite the many successes of gauge theory there is
still a variety of open problems to be solved. Some of them are connected
with the structure of the gauge orbit space. That one cannot circumvent to 
study this space was first brought to the attention of physicists by the 
discovery of what was later called the Gribov ambiguity \cite{Gribov}.
It originates from the nontrivial bundle structure of the factorization by 
local gauge transformations \cite{Singer:Gribov}. Another peculiarity is that the
gauge orbit space is a stratified space rather than a smooth manifold. 
It consists of a generic stratum and several nongeneric strata
that form singularities \cite{ArmsMarsdenMoncrief,KoRo}. 
Whereas the generic stratum was studied intensively in the early 1980s, 
leading to a
geometric understanding of the Faddev-Popov procedure
\cite{BabelonViallet:FP} and of
anomalies \cite{AtiyahSinger}, the role of nongeneric strata is not
clarified yet. There are several partial results and conjectures
\cite{Asorey:Nodes,EmmrichRoemer}, however, a systematic study has still to 
be undertaken. In the present letter we make a step in
this direction. We derive an explicit description of the particular 
stratification which is induced by the orbit types of the action of
the gauge group (assumed to be $\rmSU(n)$) on the space of
connections modulo pointed local gauge transformations. The intention of this 
is to 
provide a framework in which problems can be given a concrete formulation 
and some of the hypotheses about the role of nongeneric strata can be 
tested.

The letter is organized as follows. In Section \rref{Sprelims}, we recall
the construction of the group action to be considered. In Section 
\rref{Sstab}, we give a characterization of stabilizers in terms of
Howe subgroups. The latter are described in Section \rref{Shsg} and the 
determination 
of the set of orbit types is accomplished in Section \rref{Sot}. In Section 
\rref{Sex}, we discuss the result for base manifolds $\rmS^4$, 
$\rmS^2\times\rmS^2$, and $\CC\rmP^2$.
\section{Preliminaries}
\label{Sprelims}
Let $G$ be a compact connected Lie group, $M$ a compact connected orientable 
Riemannian manifold, and $P$ a principal $G$-bundle over $M$. 
Let $\con$ and $\calG^k$ denote the spaces of connection forms and gauge 
transformations of $P$, respectively, of Sobolev class $k$. Gauge
transformations will be viewed as $G$-space morphisms $P\rightarrow G$. For 
$2k>\dim M$, $\con$ is an affine Hilbert space and $\gau$ is a Hilbert Lie 
group and the action of $\gau$ on $\con$, given by 
$$       
A^{(g)}=g^{-1}Ag + g^{-1}\diff g \,,~~~~A\in\con,g\in\gau\,,
$$     
is smooth \cite{MitterViallet,Singer:Gribov}. 
The quotient topological space $\orb = \con/\gau$
is known as the gauge orbit space of $P$. It represents the 
space of physical degrees of freedom for any gauge theory without matter
defined on $P$. Note that
$\orb$ does not depend essentially on the technical parameter $k$, because 
using a smoothing argument one can show that for $l\leq k$, $\orb$ is open 
and dense in $\calM^l$. 

As usual for Lie group actions on manifolds, $\orb$ need not be a smooth 
manifold. One has, however, the following general construction for a group
$H$ acting on a manifold $X$. First, recall that the stabilizers
of points on the same orbit in $X$ are conjugate in $H$. Thus, 
there exists a map which assigns to each orbit in $X/H$ 
the conjugacy class of stabilizers of its representatives. It is denoted by
$\type$. The disjoint decomposition
\begin{equation}\label{Gotdecomp}
X/H = \bigcup_{\tau\in\type(X/H)} \type^{-1}(\tau)\,,
\end{equation}
is called orbit type decomposition of $X/H$. The set $\type(X/H)$
carries a natural partial ordering: $\tau\leq\tau'$ iff there exist
respective representatives $S$, $S'$ such that $S\subseteq S'$.
It was shown in \cite{KoRo} that $\otgau$ is countable and that the subsets 
$\type^{-1}(\tau)$ are manifolds. Moreover, for any $\tau\in\otgau$,    
\begin{equation}\label{Gdensitythm}
\type^{-1}(\tau) \mbox{ \it is open and dense in}
\bigcup_{\tau'\geq\tau} \type^{-1}(\tau').   
\end{equation}
These properties are condensed in the notion of 'stratification', see 
\cite{KoRo}. 
An explicit description of $\otgau$ for $G=\rmSU(n)$ and $\dim
M\leq 4$ was derived in \cite{RSV:otgauclf,RSV:otgaupo}. In this letter
an analogous result will be provided for a coarser stratification of 
$\orb$. It is obtained by viewing $\orb$ as the orbit space of another 
smooth Lie group action, constructed as follows. Let $p_\ast\in P$ be fixed.
Since, for our choice of $k$, the elements of $\gau$ are continuous, 
the Lie group homomorphism
$$
\phi^k:\gau\rightarrow G\,,~~~~g\mapsto g(p_\ast)
$$
exists. Its kernel, denoted by $\pgau$, is known as the group of pointed
local gauge transformations. It is a normal Lie subgroup and the quotient 
$\gau/\pgau$ is a Lie group isomorphic, via $\phi^k$, to $G$ 
\cite{Bourbaki:Lie}. Through this isomorphism, the residual 
action of $\gau/\pgau$ on $\conp:=\con/\pgau$ defines an action of $G$ on
$\conp$. Explicitly, for $a\in G$ and $\omega\in\conp$,
\begin{equation}\label{GGaction}
\omega^{(a)} = [A^{(g)}]_\ast\,,
\end{equation}
where $A\in\con,g\in\gau$ such that $[A]_\ast=\omega$, $\phi^k(g)=a$.
Note that the orbit space $\conp/G$ is homeomorphic to $\orb$
\cite{Bourbaki:Top}. 
It is known that $\conp$ is a smooth manifold and that $\con$ is a smooth 
locally trivial principal $\pgau$-bundle over $\conp$ \cite{MitterViallet}. 
Using local triviality, the action of $G$ on 
$\conp$ can be seen to be smooth. It was shown in \cite{KoSa} that the 
orbit type decomposition \gref{Gotdecomp} of $\orb$ induced from $G$-action
on $\conp$ is again a 
stratification and that it is encoded, in the sense of \gref{Gdensitythm}, 
in $\otG$. 

For $\omega\in\conp$ and $A\in\con$, let $G_\omega$ and $\gau_A$ denote the 
stabilizers under $G$-action and $\gau$-action, respectively. If $\omega =
[A]_\ast$, \gref{GGaction} implies
\begin{equation}\label{GGstab}
\phi^k(\gau_A)=G_\omega\,.
\end{equation}
Thus, the stratification of $\orb$ induced from $\otG$ is coarser than that
induced from $\otgau$. In particular, $\phi^k$ projects to a surjective map
\begin{equation}\label{Gotmap}
\otgau\rightarrow\otG\,.
\end{equation}
In the sequel, we are going to determine $\otG$ for $G=\rmSU(n)$ and 
$\dim M=2,3,4$.  
\section{Characterization of Stabilizers}
\label{Sstab}
Let $\omega\in\conp$ and $A\in\con$ such that $\omega=[A]_\ast$. Let $P_A$ and 
$H_A$ denote, respectively, the holonomy bundle and holonomy group of $A$, 
based at $p_\ast$. Let $\rmC_G(\cdot)$ denote the centralizer in $G$. 
According to \gref{GGstab} and the well-known relation 
$\phi^k(\gau_A)=\rmC_G(H_A)$ \cite{Fischer,NaRa},
\begin{equation}\label{Gstab}
G_\omega = \rmC_G(H_A)\,.
\end{equation}
\begin{Definition}\label{DHSG}
A subgroup $H\subseteq G$ is called Howe iff $H=\rmC_G(K)$ for
some subset $K\subseteq G$.
A (smooth) reduction of $P$ to a subgroup $H\subseteq G$ is called 
holonomy-induced iff it possesses a (smooth) connected reduction to some 
subgroup $H_0$ which obeys $\rmC_G^2(H_0)=H$.
\end{Definition}
Note that Howe subgroups can be equivalently characterized by the property 
$H=\rmC_G^2(H)$. Moreover, by definition, the structure group of a 
holonomy-induced bundle reduction is always Howe. 
\begin{Proposition}\label{Pstab}
Assume $\dim M\geq 2$. For a subgroup $S\subseteq G$, the following
assertions are equivalent:
\\
{\rm (a)} $S$ is a stabilizer of $G$-action on $\conp$,
\\
{\rm (b)} $S$ is Howe and a holonomy-induced reduction of $P$ to $\rmC_G(S)$
exists. 
\end{Proposition}
\begin{Proof}
(a) $\Rightarrow$ (b): Due to \gref{Gstab}, 
$S=\rmC_G(H_A)$ for some $A\in\con$. Hence, $S$ is Howe. For $k$ large 
enough, $A$ is of class
$C^1$, so that $P_A$ is of class $C^2$. Due to standard smoothing theory, 
$P_A$ is $C^2$-isomorphic to some 
smooth reduction $Q$ (obviously connected) of $P$ to $H_A$. The extension 
$Q\cdot\rmC_G^2(H_A)$ is a holonomy-induced reduction of $P$
to the subgroup $\rmC^2_G(H_A) = \rmC_G(S)$.
\\
(b) $\Rightarrow$ (a): By assumption, there exists a smooth connected 
reduction
$Q$ of $P$ to some subgroup $H$ obeying $\rmC^2_G(H) = \rmC_G(S)$. Since $S$
is Howe, this implies $\rmC_G(H) = S$. Being connected, $Q$ is the holonomy
bundle, and $H$ the holonomy group, of some smooth connection $A$ on $P$ 
\cite{KoNo} (this requires $\dim M\geq 2$). Then \gref{Gstab} yields
$G_{[A]_\ast} = S$. 
\qed
\end{Proof}
\begin{Remark}
(i) The subgroups $S$ and $\rmC_G(S)$ form a so-called reductive Howe 
dual pair in $G$. Such pairs play an important role in the representation 
theory of Lie groups \cite{Howe}. 

(ii) Proposition \rref{Pstab} 
implies, in particular, that $\otG$ does not depend on $k$.

(iii) While stabilizers of $G$-action on $\conp$ are characterized by the 
mere existence of certain bundle reductions, those of $\gau$-action on
$\con$ are characterized by the reductions themselves, see 
\cite{RSV:otgauclf}.
\end{Remark}
According to Proposition \rref{Pstab}, $\otG$ is a subset of $\Howe(G)$,
the set of conjugacy classes of Howe subgroups of $G$, and its partial
ordering is induced from the latter. Thus, the determination of $\otG$
proceeds through that of $\Howe(G)$.
\section{The Howe Subgroups of $\rmSU(n)$}
\label{Shsg}
General references for the determination of $\Howe(G)$ are
\cite{Przebinda,Rubenthaler,meine}. The case of $\rmSU(n)$, however, is 
simpler than the
general case. Let $\Sub_\ast(\rmM_n(\CC))$ denote the set of unital
$\ast$-subalgebras of $\rmM_n(\CC)$, the algebra of complex $(n\times
n)$-matrices, modulo conjugacy under $\rmU(n)$.
\begin{Lemma}\label{LHSG}
Intersection with $\rmSU(n)$ yields a $1$-$1$ relation between
$\Sub_\ast(\rmM_n(\CC))$ and $\Howe(\rmSU(n))$.
\end{Lemma}
\begin{Proof}
Notice that any $L\in\Sub_\ast(\rmM_n(\CC))$ is spanned by its
unitary elements which are also unitary in $\rmM_n(\CC)$. Hence
$
\rmC_{\rmM_n(\CC)} (L)
=
\rmC_{\rmM_n(\CC)} (L\cap\rmSU(n))
$.
Using this and the double commutant theorem one can establish the
1-1 relation on the level of subalgebras and subgroups. It obviously survives
the passage to conjugacy classes.
\qed
\end{Proof}
The set $\Sub_\ast(\rmM_n(\CC))$ can be described as follows.
Let $\rmK(n)$ denote the collection of pairs $J=(\bfk,\bfm)$ of sequences 
$\bfk=(k_1,\dots, k_r)$, $\bfm =(m_1,\dots, m_r)$, $r=1,2,3,\dots,n$, 
consisting of positive integers such that $\bfk \cdot\bfm = 
\sum_{i=1}^r k_im_i = n$. Any $J\in\rmK(n)$ defines a decomposition
$$
\CC^n=\left(\CC^{k_1}\otimes\CC^{m_1}\right)
\oplus\cdots\oplus
\left(\CC^{k_r}\otimes\CC^{m_r}\right)
$$
and an associated injective homomorphism
$$
\prod_{i=1}^r
\rmM_{k_i}(\CC)\rightarrow\rmM_n(\CC)
\,,~~~~
(D_1,\dots,D_r)  
\mapsto      
\bigoplus_{i=1}^r
D_i\otimes\II_{m_i}
\,.
$$
The image will be denoted by $\rmM_J(\CC)$ and its intersection with 
$\rmSU(n)$ by $\SUJ$. 
We introduce an equivalence relation
on $\rmK(n)$: $J\sim J'$ iff they differ by a
simultaneous permutation of $\bfk$ and $\bfm$. Let $\hat{\rmK}(n)$ denote 
the set of equivalence classes. As a basic fact, the map
$J\mapsto\rmM_J(\CC)$ induces a bijection from $\hat{\rmK}(n)$ onto
$\Sub_\ast(\rmM_n(\CC))$. Thus, Lemma \rref{LHSG} yields
\begin{Proposition}\label{HSG}
The map $J\mapsto\SUJ$ induces a bijection from $\hat{\rmK}(n)$ onto
$\Howe(\rmSU n)$.\qed
\end{Proposition}
\begin{Example} 
For $J=((1),(n))$ and $((n),(1))$, $\SUJ$ is the center and
the whole group, respectively. For
$J=((1,\stackrel{n}{\dots},1),(1,\stackrel{n}{\dots},1))$, $\SUJ$ is
the maximal torus of $\rmSU(n)$. For $J=((2,3),(1,1))\in\rmK(5)$, $\SUJ = 
\rmS(\rmU2\times\rmU 3)$, the symmetry group of the standard model. 
In the grand unified $\rmSU(5)$-model this is the subgroup to which
$\rmSU(5)$ is broken by the Higgs mechanism. 
For details on the structure of $\SUJ$, see \cite{RSV:otgauclf}.
\end{Example}
Next, consider the partial ordering of $\Howe(\rmSU(n))$, which obviously
coincides with that of $\Sub_\ast(\rmM_n(\CC))$. The latter can be
described in terms of inclusion matrices or Bratteli diagrams, see 
\cite[Lemma 3.1]{RSV:otgaupo}. Here we only give operations to create direct
successors. Since $\Howe(\rmSU(n))$ is finite this will allow us to
reconstruct the partial ordering and to draw Hasse diagrams. Let 
$J=(\bfk,\bfm)\in\rmK(n)$. Consider the following two operations.

{\it Splitting:} Choose $i$ such that $m_i\neq 1$ and choose positive
integers $m_{i1},m_{i2}$ such that $m_i=m_{i1}+m_{i2}$. Define
$J'=(\bfk',\bfm')$, where
\begin{eqnarray*}
\bfk' & = & (k_1,\dots,k_{i-1},k_i,k_i,k_{i+1},\dots,k_r)\,,
\\ 
\bfm' & = & (m_1,\dots,m_{i-1},m_{i1},m_{i2},m_{i+1},\dots,m_r)\,.
\end{eqnarray*}
By construction, $J'\in\rmK(n)$ and $\rmM_J(\CC)\subseteq\rmM_{J'}(\CC)$,
where 
$
\rmM_{k_i}
\subseteq
\rmM_{k_i}(\CC)\times\rmM_{k_i}(\CC)
$
diagonally and all the other factors coincide.

{\it Merging:} Choose $i<j$ such that $m_i=m_j$. Define $J'=(\bfk',\bfm')$
by
\begin{eqnarray*}
\bfk' & = &
(k_1,\dots,k_{i-1},k_i+k_j,k_{i+1},\dots,\widehat{k_j},\dots,k_r)\,,
\\
\bfm' & = & (m_1,\dots,m_{i-1},m_i,m_{i+1},\dots,\widehat{m_j},\dots,m_r)\,.
\end{eqnarray*}
Here $\widehat{\phantom{x}}$ means that the entry is omitted. Again, 
$J'\in\rmK(n)$ and $\rmM_J(\CC)\subseteq\rmM_{J'}(\CC)$, where, up to 
conjugacy, 
$
\rmM_{k_i}(\CC)\times\rmM_{k_j}(\CC)
\subseteq
\rmM_{k_i+k_j}(\CC)
$
and all other factors coincide. 

Although the following seems to be well known, the only reference the authors 
are aware of is \cite{diss}.
\begin{Proposition}\label{Pds}
Let $J\in\rmK(n)$. The labels $J'$ of the direct successors of $\SUJ$ are
obtained by applying all possible splitting and merging operations to $J$.
\qed
\end{Proposition}
\begin{Remark}
(i) When applying splitting and merging operations to $J$ one can restrict
oneself to those that yield inequivalent $J'$. 

(ii) Obviously, taking the centralizer inverts the partial ordering relation.
Thus, Proposition \rref{Pds} also yields an algorithm to create direct
predecessors. 
\end{Remark}
\begin{Example} 
Consider $\rmSU(4)$. The center has label $J=((1),(4))$.
Two splitting operations can be applied, yielding
$J'_1=((1,1),(1,3))$ and $J'_2 = ((1,1),(2,2))$. At the next stage,
a splitting operation can be applied to $J'_1$, yielding
$J^{\prime\prime}_1 = ((1,1,1),(1,1,2))$. Two splitting
operations can be applied to $J'_2$. Their results are equivalent to
$J^{\prime\prime}_1$. This means that $\rmSU(J'_1)$ and $\rmSU(J'_2)$ have
common direct successor $\rmSU(J^{\prime\prime}_1)$. Furthermore, a merging
operation can be applied to $J'_2$, yielding $((2),(2))$. Continuing
the procedure one can easily construct the Hasse diagram of $\Howe(\rmSU(n))$
for $n=4$ and, similarly, for any other value of $n$. The results for
$n=2,\dots,5$ are shown in Figure \rref{F} at the end. Note that the diagrams 
are symmetric w.r.t. reflection at the vertical central axis and simultaneous
interchange of $\bfk$ and $\bfm$. Of course, this is due to Remark
(ii) above.
\end{Example}
\section{The Set of Orbit Types}
\label{Sot}
The following lemma was proved in \cite[Thm.~6.2]{RSV:otgauclf}. 
\begin{Lemma}\label{Lholind}
Any reduction of a principal $\rmSU(n)$-bundle to a Howe subgroup is
holonomy-induced.
\qed
\end{Lemma}
\begin{Remark} 
Lemma \rref{Lholind} does not hold, for example, for $\rmSO(n)$. 
\end{Remark}
\begin{Lemma}\label{Lctr}
For $J=(\bfk,\bfm)$, $\rmC_{\rmSU(n)}(\SUJ)$ is conjugate to $\SUJc$, where
$J^c=(\bfm,\bfk)$. \qed
\end{Lemma}
For the reductions 
of $P$ to $\SUJc$, the following classification was derived in 
\cite{RSV:otgauclf}. Let the symbol $\gcd{~}$ denote
the greatest common divisor of the integers enclosed. Define integers 
$\twk_i$ by $\twk_i\gcd{\bfk} = k_i$, $i=1,\dots,r$.
Let $H^\rmeven(M,\ZZ)$ denote the even degree part of the integral cohomology 
ring
$H^\ast(M,\ZZ)$. We introduce the notation
$$
H^{(\bfm)} (M,\ZZ) = \{\alpha\in\prod_{i=1}^r H^\rmeven(M,\ZZ) 
~|~
\alpha_i^{(0)}=1, \alpha_i^{(2j)} = 0 \mbox{ for } j>m_i \}\,.
$$
Here $\alpha=(\alpha_1,\dots,\alpha_r)$ and $\alpha_i^{(2j)}$ denotes the 
component of $\alpha_i$ of degree $2j$.
Note that each of the $\alpha_i$ can be viewed as the (total) Chern class of
a $\rmU(m_i)$-bundle over $M$. Finally, let $c(P)$ denote the (total) Chern 
class of $P$ and $\beta_{\gcd{\bfk}}:H^1(M,\ZZ_{\gcd{\bfk}})\rightarrow
H^2(M,\ZZ)$ the Bockstein homomorphism associated to the short exact sequence
$0\rightarrow\ZZ\rightarrow\ZZ\rightarrow\ZZ_{\gcd{\bfk}}\rightarrow 0$
of coefficient homomorphisms. Consider the system of equations
\begin{eqnarray}\label{GKJM}
\sum_{i=1}^r \twk_i\alpha_i^{(2)} & = & \beta_{\gcd{\bfk}}(\xi)\,,
\\ \label{GKJP}
\alpha_1^{k_1}\cdots\alpha_r^{k_r} & = & c(P)
\end{eqnarray}
in the indeterminates $\alpha\in H^{(\bfm)}(M,\ZZ)$ and $\xi\in
H^1(M,\ZZ_{\gcd{\bfk}})$. The following was stated in
\cite{RSV:otgauclf} as Theorem 5.16.
\begin{Lemma}\label{Lbunred}
Assume $\dim M\leq 4$. The reductions of $P$ to the subgroup $\SUJc$ are
classified, up to isomorphy, by the solutions of \gref{GKJM}, \gref{GKJP}. 
\qed
\end{Lemma}
\begin{Remark} 
Eq.~\gref{GKJM} is a relation between the characteristic classes which classify principal $\SUJc$-bundles. It emerges
from their construction. Eq.~\gref{GKJP} represents the
condition that the $\SUJc$-bundle labelled by $\alpha$, $\xi$ is a
reduction of $P$.
\end{Remark}
Eq.~\gref{GKJP} actually contains two equations, sorted by degree,
{\arraycolsep2pt
\begin{eqnarray}\label{GKJP2}
\sum_{i=1}^r k_i \alpha_i^{(2)}
& = &
0\,,
\\ \label{GKJP4}
\sum_{i=1}^r k_i\alpha_i^{(4)}
+ \sum_{i=1}^r\frac{k_i^2-k_i}{2}
\left(\alpha_i^{(2)}\right)^2
+\!\!\! \sum_{1\leq i<j\leq r} \!\!k_ik_j
\alpha_i^{(2)}\alpha_j^{(2)}
& = &
c_2(P),~~~~~
\end{eqnarray}
}

\noindent
where $c_2(P)$ denotes the second Chern class of $P$. Note that \gref{GKJP2}
already follows from \gref{GKJM}, because $\gcd{\bfk}\beta_{\gcd{\bfk}} =
0$. For our purposes, it suffices to know whether the system \gref{GKJM},
\gref{GKJP} has a solution or not. Let $H^{(\bfm)}_F(M,\ZZ)$ denote the
torsion-free part of $H^{(\bfm)}(M,\ZZ)$.
\begin{Lemma}\label{Leq}
The system of Eqs. \gref{GKJM}, \gref{GKJP} possesses a solution if and
only if Eq. \gref{GKJP} possesses a solution $\alpha\in H^{(\bfm)}_F(M,\ZZ)$. 
\end{Lemma}
\begin{Proof} 
Let $\alpha\in H^{(\bfm)}_F(M,\ZZ)$ be a solution of \gref{GKJP}. Then, due to 
\gref{GKJP2}, choosing $\xi=0$ yields a solution to \gref{GKJM}. Conversely, 
let $\alpha$, $\xi$ be a solution of \gref{GKJM} and \gref{GKJP}. 
Decompose $\alpha=\alpha_T+\alpha_F$ into torsion and torsion-free part. 
Eq.~\gref{GKJP2} is satisfied by $\alpha_T$ and $\alpha_F$ 
independently. By orientability of $M$, $\alpha_T$ does not contribute to 
\gref{GKJP4}. It follows that $\alpha_F$ solves \gref{GKJP}.
\qed
\end{Proof}
Let $\rmK(P)$ denote the subset of $\rmK(n)$ of elements $J=(\bfk,\bfm)$ for 
which \gref{GKJP} possesses a solution in $H^{(\bfm)}_F(M,\ZZ)$. Since 
simultaneous permutations of $\bfk$ and $\bfm$ do
not affect this property, we can pass to  the set of equivalence classes,
which will be denoted by 
$\hat{\rmK}(P)$.
\begin{Theorem}
Assume $\dim M=2,3,4$. Then the map $J\mapsto\SUJ$ induces a bijection from 
$\hat{\rmK}(P)$ onto $\otG$.
\end{Theorem}
\begin{Proof}
Let $J\in\rmK(n)$. 
It suffices to check that $J\in\rmK(P)$ iff $\SUJ$ is a stabilizer.
Due to Proposition \rref{Pstab} and Lemma \rref{Lctr}, $\SUJ$ is a stabilizer 
iff $P$ admits a 
holonomy-induced reduction to $\SUJc$. According to Lemma \rref{Lholind}, one 
can omit holonomy-induced here. Then the assertion follows from Lemmas 
\rref{Lbunred} and \rref{Leq}.
\qed
\end{Proof}
As a result, the determination of $\otG$ is reduced to a discussion of the
solvability of the system of equations \gref{GKJP2}, \gref{GKJP4}. Let us 
remark that, contrary to that, the elements of $\otgau$ are characterized by 
the solutions $\alpha,\xi$ themselves (cf.~Lemma \rref{Lbunred} and Remark 
(iii) after Proposition \rref{Pstab}). On the level of 
the data $J,\alpha,\xi$, the map \gref{Gotmap} reads $(J,\alpha,\xi)\mapsto J$. 
\section{Examples}
\label{Sex}
In dimensions $2,3$, any principal $\rmSU(n)$-bundle is trivial, hence 
can be reduced to any of the subgroups $\SUJ$. Therefore,
$\otG=\Howe(\rmSU(n))$.
In dimension $4$, Eq.~\gref{GKJP4} may give rise to a linear, 
bilinear, or quadratic Diophantine equation. Examples for these 3 types
are provided by $M=\rmS^4$, $\rmS^2\times\rmS^2$, and $\CC\rmP^2$,
respectively. For all of them, $H^{(\bfm)}(M,\ZZ) = H^{(\bfm)}_F(M,\ZZ)$.
\subsection{Base Manifold $M=\rmS^4$}
Since $H^2(\rmS^4,\ZZ)=0$, Eq.~\gref{GKJP2} is trivially satisfied. 
We parametrize $c_2(P) = c_P\gamma$ and $\alpha_i^{(4)} = b_i\gamma$,
$i=1,\dots,r$, where $\gamma$ is a generator of $H^4(\rmS^4,\ZZ)$.
Eq.~\gref{GKJP4} yields the linear Diophantine equation
\begin{equation} \label{Glin}
\sum_{i=1}^r k_i b_i=c_P\,.
\end{equation}
Recall that $b_i\in\ZZ$ if $m_i\neq 1$ and $b_i=0$ otherwise. Thus,
\gref{Glin} has a solution iff $c_P$ is a multiple of $\gcd{\bfkred}$,
where $\bfkred$ is obtained from $\bfk$ by deleting all members $k_i$ for
which $m_i=1$. The case $\bfkred=\emptyset$ can be consistently incorporated
by putting $\gcd{\emptyset}=0$. Denoting $\gcd{\bfkred}$ by
$\rmd_{\rmS^4}(J)$, 
\begin{equation}\label{GS4K(P)}
\rmK(P) = \{J\in\rmK(n)~|~\rmd_{\rmS^4}(J)\mbox{ divides }c_P\}\,.
\end{equation}
\subsection{Base Manifold $M=\rmS^2\times\rmS^2$}
\newcommand{\x}{\!\times\!}
Let $\gamma$ be a generator of $H^2(\rmS^2,\ZZ)$. Then $H^2(M,\ZZ)$ and
$H^4(M,\ZZ)$ are generated by $\gamma\x 1, 1\x\gamma$ and 
$\gamma\x\gamma$, respectively. We expand $\alpha_i^{(2)} =
a_{1i}~\gamma\x 1 + a_{2i}~1\x\gamma$ and $c_2(P) = c_P~
\gamma\x\gamma$. Eqs.~\gref{GKJP2}, \gref{GKJP4} read
\begin{eqnarray}\label{GS2KJP2}
\sum_{i=1}^rk_ia_{li}
& = &
0\,,~~~~l=1,2\,,
\\ \label{GS2KJP4}
\gcd{\bfkred}b + \sum_{i,j=1}^r k_i(k_j-\delta_{ij})a_{1i}a_{2j}
& = &
c_P\,.
\end{eqnarray}
Here $b\in\ZZ$ is an indeterminate (thus we made use of our previous result)
and $\delta_{ij}$ denotes the Kronecker symbol.
Since the case $r=1$ is trivial, we may assume $r\geq 2$. Then the set of
solutions of Eq.~\gref{GS2KJP2} can be parametrized by integers
$t_{l,pq}$, $1 \leq p < q \leq r$, as follows \cite{Skolem:Diophant}:
$$
a_{li} = -\sum_{m=1}^{i-1} \twk_m t_{l,mi} + \sum_{m=i+1}^r \twk_m t_{l,im}
\,,~~~~i=1,\dots,r, l=1,2\,.
$$
Unless $r=2$, the parametrization is not $1$-$1$, but it generates all
solutions, which suffices for our purposes. Insertion into
\gref{GS2KJP4} yields the bilinear Diophantine equation
\begin{equation} \label{GS2eq}
\gcd{\bfkred}b
+
\sum_{1\leq m<i\leq r\atop 1\leq n<j\leq r}
L_{mi,nj}t_{1,mi}t_{2,nj}
=
c_P\,,
\end{equation}
where
$$
L_{mi,nj}
=
\gcd{\bfk} ~\twk_m \twk_i~
(\twk_n (\delta_{mj} - \delta_{ij}) +  \twk_j (\delta_{ni} - \delta_{mn}))\,.
$$
It is well known that a bilinear form over $\ZZ$ can take as value 
any multiple of the greatest common divisor of its coefficients 
\cite{Skolem:Diophant}. Denote the latter by $\gcd{L}$. Then 
\gref{GS2eq} has a solution iff $c_P$ 
is a multiple of $d_{\rmS^2\x\rmS^2}(J) = \gcd{\gcd{\bfkred},\gcd{L}}$. 
The case $r=1$ can be consistently incorporated by setting 
$L=\emptyset$. Thus,
\begin{equation}\label{GS2K(P)}
\rmK(P) = \{J\in\rmK(n) ~|~ d_{\rmS^2\x\rmS^2}(J)\mbox{ divides }c_P\}\,.
\end{equation}
As $d_{\rmS^2\x\rmS^2}(J)$ divides $d_{\rmS^4}(J)$, there are
'more' orbit types over $\rmS^2\x\rmS^2$ than over $\rmS^4$. 
Note that \gref{GS2K(P)} holds for $M=\rmT^4$ (the $4$-torus), too.

To compute $\gcd{L}$, observe that, besides $0$ and up to a sign, there are
two types of coefficients, namely,
\begin{eqnarray*}
L_{mi,mj} & = & -\gcd{\bfk} \twk_m \twk_i\twk_j
\,,~~~1\leq m<i<j\leq r
\\
L_{mi,mi} & = & -\gcd{\bfk} \twk_m \twk_i (\twk_m + \twk_i)
\,,~~~1\leq m<i\leq r\,.
\end{eqnarray*}
Thus, $\gcd{L}$ is just the greatest common divisor of
all these numbers. 
As an example, consider $J = ((4,4,6),(1,1,2))\in\rmK(20)$. Here
$\gcd{\bfk}=2$ and $\gcd{\bfkred}=6$. The relevant coefficients are
$2\twk_1\twk_2\twk_3=2^3\cdot 3$,
$2\twk_1\twk_2(\twk_1+\twk_2) = 2^5$, and
$2\twk_1\twk_3(\twk_1+\twk_3) = 2^2\cdot 3\cdot 5$. They yield $\gcd{L}=4$. 
Hence, $d_{\rmS^2\x\rmS^2}(J) = 2$.

The numbers $d_{\rmS^4}(J)$ and $d_{\rmS^2\x\rmS^2}(J)$ for $J\in\rmK(n)$,
$n=2,\dots,5$, are given in Figure \gref{F}. Using \gref{GS4K(P)} and
\gref{GS2K(P)}, the respective Hasse diagrams of $\otG$ can be read off
directly from this figure. 
\subsection{Base Manifold $M=\CC\rmP^2$}
Let $\gamma$ be a generator of $H^2(\CC\rmP^2,\ZZ)$. Then
$H^4(\CC\rmP^2,\ZZ)$ is generated by $\gamma^2$. With
$\alpha_i^{(2)} = a_i~\gamma$ and $c_2(P) = c_P~\gamma^2$,
Eqs.~\gref{GKJP2} and \gref{GKJP4} read
\begin{eqnarray}\nonumber
\sum_{i=1}^r~k_i a_i & = & 0\,,
\\ \label{GCP2KJP4}
\gcd{\bfkred}b + \frac{1}{2}\sum_{i,j=1}^r k_i(k_j-\delta_{ij})a_ia_j
& = &
c_P\,,
\end{eqnarray}
where $b\in\ZZ$ is an indeterminate. Notice that all the coefficients are
integral. Thus, here we are facing a quadratic Diophantine equation or,
phrased differently, the representation problem for a quadratic form over
$\ZZ$. Since a profound discussion, apart from some simple examples,
requires methods from number theory (which is beyond the scope of this
letter), it will be given elsewhere. Let us only consider the cases $n=2$
and $n=3$. For $n=2$, only $J=((1,1),(1,1))$ needs to be considered. By
eliminating $a_2$, \gref{GCP2KJP4} becomes $-a_1^2=c_P$. Thus, the orbit
type labelled by $J$ is present iff $-c_P$ is a square. Next, consider $n=3$.
Type $J=((1,1),(1,2))$ is always present, because here $\gcd{\bfkred}=1$.
For type $J=((2,1),(1,1))$, \gref{GCP2KJP4} becomes
$-3a_1^2=c_P$. Thus, this orbit type is present iff $-c_P$ is $3$ times a
square. Finally, for $J=((1,1,1),(1,1,1))$, after elimination of $a_3$,
\gref{GCP2KJP4} reads
\begin{equation} \label{GCP2eq}
-(a_1^2+a_1a_2+a_2^2) = c_P\,.
\end{equation}
Here it is no longer
obvious for which $c_P$ this equation has a solution. Of course, since the
l.h.s.~is negative definite, for any given $c_P$, only finitely many values
of $a_1,a_2$ have to be tested, so one could use the help of a computer. In
fact, for some $J$ this might be the only way to solve the problem. For
\gref{GCP2eq}, however, more elaborate arguments \cite{Jones} show that it is
solvable, and hence type $J$ is present in the gauge orbit space, iff
$c_P\leq 0$ and 

(i) $c_P\neq -3^m(3n+2)$, $\forall~m,n\in\ZZ\,,~~m\geq 0$.

(ii) in the decomposition of $-c_P$ into prime factors, any prime with $p=5$ 
or $11$ $\mod 12$ appears to an even power.

Thus, a solution exists for $-c_P=1$, $3$, $4$, $7$, $9$, $12$ etc., but not 
for $2$, $5$, $6$, $8$, $10$, $11$ etc. . While (i) determines an arithmetic 
progression of bundles $P$ for which type $J$ is not present, (ii) picks out
additional such $P$ in a sporadic manner. Note that, while (ii) is a 
condition whose form is peculiar to binary quadratic forms, (i) is an
analogue of the condition appearing in the famous result of Gau{\ss} that a 
positive 
integer is a sum of $3$ squares iff it is not of the form $4^m(8n+7)$.
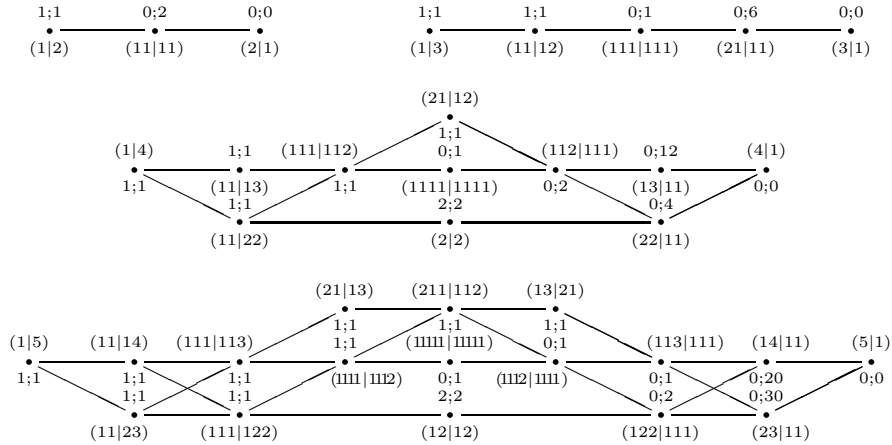
\begin{figure}
\caption{Hasse diagrams of $\Howe(\rmSU(n))$ for $n=2,\dots,5$. All edges
are directed from left to right. Vertices are labelled by $J=(\bfk,\bfm)$,
written as $(k_1k_2\dots|m_1m_2\dots)$ omitting commata, and numbers 
$\facs{d_{\rmS^4}(J)}{d_{\rmS^2\x\rmS^2}(J)}$.\label{F}}
\begin{center}
\unitlength1.4cm
\begin{picture}(2,0.8)
\put(0,0.5){
\plene{0,0}{tc}{(1|2)}
\plene{0,0}{bc}{\facs{1}{1}}
\plene{1,0}{tc}{(11|11)}
\plene{1,0}{bc}{\facs{0}{2}}
\whole{2,0}{tc}{(2|1)}
\whole{2,0}{bc}{\facs{0}{0}}
}
\end{picture}
\hspace{2cm}
\begin{picture}(4,0.8)
\put(0,0.5){
\plene{0,0}{tc}{(1|3)}
\plene{0,0}{bc}{\facs{1}{1}} 
\plene{1,0}{tc}{(11|12)}
\plene{1,0}{bc}{\facs{1}{1}}
\plene{2,0}{tc}{(111|111)}
\plene{2,0}{bc}{\facs{0}{1}}
\plene{3,0}{tc}{(21|11)}
\plene{3,0}{bc}{\facs{0}{6}}
\whole{4,0}{tc}{(3|1)}
\whole{4,0}{bc}{\facs{0}{0}}
}
\end{picture}
\\
\begin{picture}(6,1.8)
\put(0,1){
\plene{0,0}{bc}{(1|4)}
\plzmee{0,0}{tc}{\facs{1}{1}}
\plene{1,0}{tc}{(11|13)}
\plene{1,0}{bc}{\facs{1}{1}} 
\plzee{1,-0.5}{tc}{(11|22)}    
\plenz{1,-0.5}{bc}{\facs{1}{1}}
\plzee{2,0}{br}{(111|112)\!\!\!\!\!\!}
\plene{2,0}{tc}{\facs{1}{1}}
\plzmee{3,0.5}{bc}{(21|12)} 
\plzmee{3,0.5}{tc}{\facs{1}{1}}
\plene{3,0}{tc}{(1111|1111)}
\plene{3,0}{bc}{\facs{0}{1}}
\plenz{3,-0.5}{tc}{(2|2)}
\plenz{3,-0.5}{bc}{\facs{2}{2}}
\plene{4,0}{bl}{\!\!\!\!\!\!(112|111)}
\plzmee{4,0}{tc}{\facs{0}{2}}
\plene{5,0}{tc}{(13|11)} 
\plene{5,0}{bc}{\facs{0}{12}}
\plzee{5,-0.5}{tc}{(22|11)}
\plzee{5,-0.5}{bc}{\facs{0}{4}}    
\whole{6,0}{bc}{(4|1)}
\whole{6,0}{tc}{\facs{0}{0}}
}
\end{picture}
\\
\begin{picture}(8,1.8)
\put(0,1){
\plene{0,0}{bc}{(1|5)}
\plzmee{0,0}{tc}{\facs{1}{1}}
\plene{1,0}{br}{(11|14)\!\!\!\!\!\!}
\plzmee{1,0}{tc}{\facs{1}{1}}
\plzee{1,-0.5}{tr}{(11|23)\!\!\!\!\!\!}
\plene{1,-0.5}{bc}{\facs{1}{1}}                
\plzee{2,0}{br}{(111|113)\!\!\!\!\!\!}
\plene{2,0}{tc}{\facs{1}{1}}
\plzee{2,-0.5}{tc}{(111|122)}
\plenz{2,-0.5}{bc}{\facs{1}{1}}
\plene{3,0.5}{bc}{(21|13)}    
\plene{3,0.5}{tc}{\facs{1}{1}}
\plzee{3,0}{tl}{\!\!\!\!\!\!(\!1\!1\!1\!1|1\!1\!1\!2)} 
\plene{3,0}{bc}{\facs{1}{1}}
\plene{4,0.5}{bc}{(211|112)}
\plzmee{4,0.5}{tc}{\facs{1}{1}}
\plene{4,0}{bc}{(1\!1\!1\!1\!1|1\!1\!1\!1\!1)}
\plene{4,0}{tc}{\facs{0}{1}}
\plenz{4,-0.5}{tc}{(12|12)}
\plenz{4,-0.5}{bc}{\facs{2}{2}} 
\plzmee{5,0.5}{bc}{(13|21)}    
\plzmee{5,0.5}{tc}{\facs{1}{1}}    
\plene{5,0}{tr}{(\!1\!1\!1\!2|1\!1\!1\!1)\!\!\!\!\!\!}
\plzmee{5,0}{bc}{\facs{0}{1}}
\plene{6,0}{bl}{\!\!\!\!\!\!(113|111)}
\plzmee{6,0}{tc}{\facs{0}{1}}
\plzee{6,-0.5}{tc}{(122|111)}
\plene{6,-0.5}{bc}{\facs{0}{2}}
\plene{7,0}{bl}{\!\!\!\!\!\!(14|11)}
\plene{7,0}{tc}{\facs{0}{20}}    
\plzee{7,-0.5}{tl}{\!\!\!\!\!\!(23|11)}
\plzee{7,-0.5}{bc}{\facs{0}{30}} 
\whole{8,0}{bc}{(5|1)}
\whole{8,0}{tc}{\facs{0}{0}}
}
\end{picture}
\end{center}
\end{figure}
\section*{Acknowledgements}
We would like to thank I.P.~Volobuev for sharing his insights.
\end{article}

\begin{thebibliography}{99}
%
\bibitem{ArmsMarsdenMoncrief}
Arms, J.M., Marsden, J.E., Moncrief, V.: {\it Commun.~Math.~Phys.\/} 
{\bf 78} (1981) 455
%
\bibitem{Asorey:Nodes}
Asorey, M., Falceto, F., L\'opez, J.L., Luz\'on, G.:
{\it Phys.~Lett.} {\bf B 345} (1995) 125
%
\bibitem{AtiyahSinger}
Atiyah, M.F. and Singer, I.M.:
{\it Proc.~Nat.~Acad.~Sci.~USA} {\bf 81} (1984) 2597
%
\bibitem{BabelonViallet:FP}
Babelon, O. and Viallet, C.-M.: {\it Phys. Lett.} {\bf B 85} (1979) 246
%
\bibitem{Bourbaki:Lie}
Bourbaki, N.:
{\it Lie Groups and Lie Algebras}, Springer, 1989,
Prop.~III.3.28
%
\bibitem{Bourbaki:Top}
Bourbaki, N.:
{\it General Topology}, Springer, 1989,
Prop.~III.2.22
%
\bibitem{EmmrichRoemer}
Emmrich, C. and R{\"o}mer, H.:
{\it Commun.~Math.~Phys.\/} {\bf 129} (1990) 69
%
\bibitem{Fischer}
Fischer, A.E.:
{\it Commun.~Math.~Phys.\/} {\bf 113} (1987) 231
%
\bibitem{Gribov}
Gribov, V.N.:
{\it Nucl.~Phys.\/}, {\bf B 139} (1978) 1
%
\bibitem{Howe}
Howe, R.:
{\it Proc.~Symp.~Pure Math.\/} {\bf 33}, part 1, (1979) 275
%
\bibitem{Jones} 
Jones, B.W.:
{\it The Arithmetic Theory of Quadratic Forms}, Math.~Assoc.~of
America, 1967
%
\bibitem{KoNo}
Kobayashi, S. and Nomizu, K.:
{\it Foundations of Differential Geometry},
Wiley Interscience, New York, 1963, ch.~II, Thm.~8.2
%
\bibitem{KoRo}
Kondracki, W. and Rogulski, J.:
{\it Dissertationes Mathematicae\/} {\bf 250},
Panstwowe Wydawnictwo Naukowe, Warszawa, 1986
%
\bibitem{KoSa}
Kondracki, W. and Sadowski, P.:
{\it J.~Geom.~Phys.\/} {\bf 3} (1986) 421
%
\bibitem{MitterViallet}
Mitter, P.K. and Viallet, C.-M.:
{\it Commun.~Math.~Phys.\/} {\bf 79} (1981) 457
%
\bibitem{NaRa}
Narasimhan, M.S. and Ramadas, T.R.:
{\it Commun.~Math.~Phys.\/} {\bf 67} (1979) 121
%
\bibitem{Przebinda}
Przebinda, T.: 
{\it J.~Funct.~Anal.\/} {\bf 81} (1988) 160
%
\bibitem{Rubenthaler}
Rubenthaler, H.:
{\it Les Paires Duales dans les Alg\`ebres de Lie R\'eductives},
Ast\'erisque {\bf 219}, Soc.~Math.~de France 1994
%
\bibitem{RSV:otgauclf} Rudolph, G., Schmidt, M. and Volobuev, I.P.:
math-ph/0003044,
%
\bibitem{RSV:otgaupo} Rudolph, G., Schmidt, M. and Volobuev, I.P.:
math-ph/0009018,
%
\bibitem{meine}
Schmidt, M.: {\it J.~Geom.~Phys.\/} {\bf 29} (1999) 283
%
\bibitem{diss}
Schmidt, M.:
{\it Gauge Orbit Types for Theories with Gauge Group $\rmSU(n)$},
Thesis, Univ.~of Leipzig, 2000
%
\bibitem{Singer:Gribov}
Singer, I.M.: 
{\it Commun.~Math.~Phys.\/} {\bf 60} (1978) 7
%
\bibitem{Skolem:Diophant}
Skolem, T.:
{\it Diophantische Gleichungen\/},
Springer, Berlin, 1938, ch.~I
%
%
%
\end{thebibliography}
\end{document}